# Analysis on rotational Doppler Effect based on modal expansion method

Hailong Zhou,[1] Jianji Dong,[1,*] Pei Zhang,[2] and Xinliang Zhang[1]

[1]Wuhan National Laboratory for Optoelectronics, School of Optoelectronic Science and Engineering, Huazhong University of Science and Technology, Wuhan, China, 430074

[2]MOE Key Laboratory for Nonequilibrium Synthesis and Modulation of Condensed Matter, Department of Applied Physics, Xi'an Jiaotong University, Xi'an 710049, China

*Corresponding author: jjdong@mail.hust.edu.cn

We theoretically investigate the optical rotational Doppler Effect using modal expansion method. We find that the frequency shift content is only determined by the surface of spinning object and the reduced Doppler shift of $(l-m)\Omega/2\pi$ is linear to the change of mode index, where $l$ is mode index of the incident orbital angular momentum (OAM) light and $m$ is the one of the OAM light reflected or transmitted from a surface rotating at a fixed speed $\Omega$. The theoretical model makes us better understand the physical processes of rotational Doppler Effect. It can provide theoretical guidance for many related applications, such as detection of rotating bodies, detection of OAM and frequency shift.

Linear Doppler Effect is a well-known phenomenon by which the frequency of a wave is shifted according to the relative velocity of the source and the observer. This frequency shift scales with both the unshifted frequency and the linear velocity, and it is extensively used in Doppler velocimetry to detect the translational motion of surfaces and fluids [1, 2]. The Doppler velocimetry is a very mature technology and has been fully explored many years ago. In recent years, there has been increasing interest in another type of Doppler Effect, namely rotational Doppler Effect, in which a spinning object with an optically rough surface may induce a Doppler shift in light reflected parallel to the rotation axis, provided that the light carries orbital angular momentum (OAM) [3-5]. The OAM light comprises a transverse angular phase profile equal to $\exp(il\theta)$, where $\theta$ is the angular coordinate and $l$ is the azimuthal index, defining the topological charge (TC) of the OAM modes [6], namely the OAM mode index. These beams have an OAM of $l\hbar$ per photon ($\hbar$ is Planck's constant divided by $2\pi$) and consist of a ring of intensity with a null at the center. OAM light has been widely used in a variety of interesting applications, such as in optical microscopy [7], micromanipulation [8-11], quantum information [12, 13], optical communication [14-16]. And it has also been applied in probing the angular velocity of spinning microparticles or objects based on rotational Doppler Effect [3-5, 17, 18].

In 2013, Padgett's group recognized that the well-known Doppler shift and Doppler velocimetry had an angular equivalent. And other works about measurement of transverse velocity based on rotational Doppler Effect were also presented [3, 5, 18-21]. In 2014, Padgett's group theoretically presented the physical mechanism of rotational Doppler Effect [22]. But this work was the case where the observer spun relative to the beam axis. A more general case is that a fixed input light illuminates a spinning object and the scattered light has a frequency shift. The mechanism of the frequency shift is still not very clear. For example, it is still not clear how the surface of object influences the frequency shift content and what is the relationship between the incident light, scattered light and the surface.

In this letter, we theoretically investigate the optical rotational Doppler Effect using modal expansion method. We find that the frequency shift content is only determined by the surface of spinning object and the reduced Doppler shift of $(l-m)\Omega/2\pi$ is linear to the change of mode index, where $l$ is the mode index of the incident orbital angular momentum (OAM) light and $m$ is the one of the OAM light reflected or transmitted from a surface rotating at a fixed speed of $\Omega$. The theoretical model makes us better understand the physical processes of rotational Doppler Effect. An example of inputting two OAM modes is presented. The model can provide theoretical guidance for many related applications, such as detection of rotating bodies, detection of OAM and frequency shift.

The rotational Doppler Effect can be regarded as an aggregative effect caused by the synchronized Linear Doppler Effect at every position of the spinning object. A laser Doppler velocimeter works based on the principle that light scattered from a moving object is frequency shifted with respect to the incident light. If a collimated beam of light of wavelength $\lambda$ (frequency $f$) is incident on a moving surface, the reduced Doppler shift is given by [2]

$$\Delta f = (\vec{d_1} - \vec{d_s})\cdot\vec{v}\lambda = (\sin\alpha - \sin\beta)vf/c, \qquad (1)$$

where $\vec{d_s}$ and $\vec{d_1}$ are the unit vectors of the incident and scattered light, $\alpha$ and $\beta$ are the incident angle and reflected angle. $\vec{v}$ is the velocity vector of the moving surface, $v = |\vec{v}|$ and $c$ is the speed of light in vacuum, shown as Fig. 1.

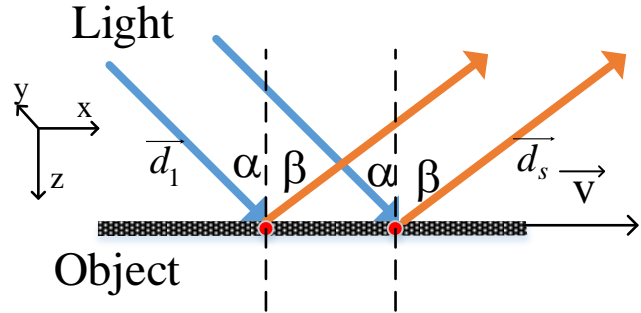

Fig. 1 Principle of Linear Doppler Effect.

When an OAM light illuminates a spinning object with a rotating speed $\Omega$, as shown in Fig. 2, we choose the mode indexes of the incident and scattered light as $l$ and $m$ respectively. The scattered light can be selected as the reflected light from the object or the transmitted light through the object. Here, the coordinate systems are always the same for the incident and scattered light, whether the scattered light is reflected light or transmitted light. This is done to avoid changing the sign of mode indexes caused by the reflection. Without loss of generality, the following analysis is also applicable to the transmitted light. In a helically phased beam with $TC = l$, the skew angle between the Poynting vector and the beam axis is $l\lambda/2\pi r$, where $r$ is the radius from the beam axis [4, 5, 23]. So the incident angle and scattered angle are $\alpha = l\lambda/2\pi r$ and $\beta = m\lambda/2\pi r$. For a well-collimated beam, the radial component of Poynting vector of an OAM mode is close to null [23], so the vectors $\vec{d_s}$, $\vec{d_1}$ and $\vec{v}$ ($|\vec{v}| = \Omega r$) are in the same plane as shown in Fig. 2(b). From Eq. (1), we can know that the reduced Doppler shift at every position of the spinning object is

$$\Delta f = v(\sin\alpha - \sin\beta)f/c \approx (l-m)\Omega/2\pi. \qquad (2)$$

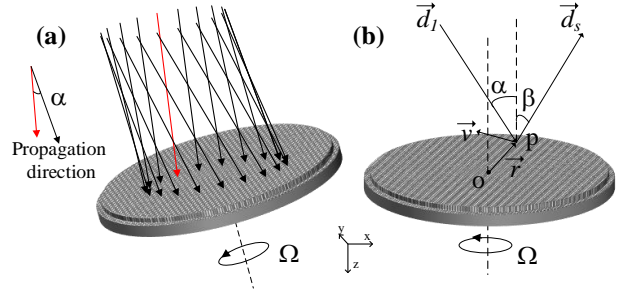

Fig. 2 (a) Schematic diagram and (b) analysis diagram of rotational Doppler Effect.

Here, the approximation $\sin\alpha \approx \alpha, \sin\beta \approx \beta$ is used. So when an OAM light illuminates a spinning object with a rotating speed $\Omega$, the scattered light has a reduced frequency shift $(l-m)\Omega/2\pi$, which is related to the mode indexes of the incident and scattered light, and the rotating speed. Eq. (2) is based on the fact that the surface is rough enough so that it can scatter the incident light in a mass of directions, i.e., the scattered light contains plentiful OAM modes. Although the rotational Doppler Effect is demonstrated from Eq. (2), it is still not clear how the surface of object influences the frequency shift content and the OAM modes.

In the following, we will make a detail analysis to the relationship between the incident light, scattered light and the spinning surface by using modal expansion method. The roughness of surface means that the scattered light contains plentiful OAM modes, namely, a pure OAM mode can be mapped into a mass of OAM modes, which are determined by the rough surface. So the spinning surface can be regarded as a mode mapping device. As shown in Fig. 3, we assume that we only consider the OAM modes with TC ranging from $l_1$ to $l_N$. The transmission matrix formula can be written as

$$\begin{pmatrix} E_{o,l_1} \\ E_{o,l_2} \\ \vdots \\ E_{o,l_N} \end{pmatrix} = \begin{pmatrix} a_{l_1 l_1} & \cdots & a_{l_N l_1} \\ \vdots & \ddots & \vdots \\ a_{l_1 l_N} & \cdots & a_{l_N l_N} \end{pmatrix} \begin{pmatrix} E_{i,l_1} \\ E_{i,l_2} \\ \vdots \\ E_{i,l_N} \end{pmatrix}, \qquad (3)$$

where $E_{i,l_p}$ $(p=1,2,..,N)$ is the light field of input OAM mode with TC=$l_p$, $E_{o,l_p}$ $(p=1,2,..,N)$ is the light field of output OAM mode with TC=$l_p$ and $a_{l_p l_q}$ is the mode mapping coefficient from OAM mode with TC=$l_p$ to OAM mode with TC=$l_q$. For a motionless surface, it is obvious that the mode mapping coefficients are constant in time. And the coefficients become periodic when the surface rotates at a fixed speed, because the cases always remain the same when the surface rotates every circle.

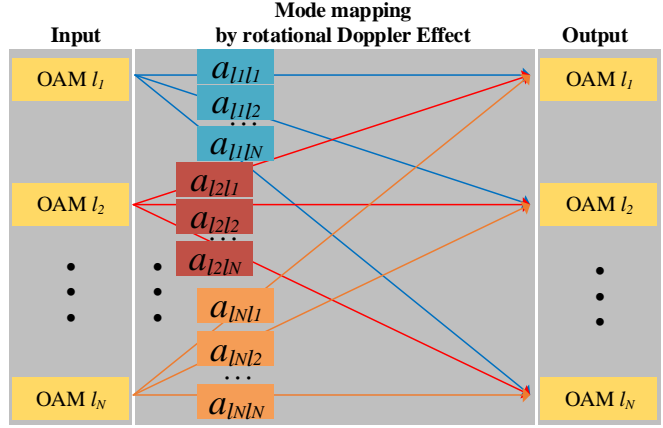

Fig. 3. Mode mapping by rotational Doppler Effect.

As we know, OAM mode conversion can be implemented by a spiral phase plate (SPP), which can add a helical phase $\exp(in\theta)$ on the illumination light [24, 25]. As shown in Fig. 4(a), when an input OAM light with TC equal to $l$ illuminates on the SPP, the TC of the output OAM mode is changed to $l+n$. The similar principle can be applied to the rotational Doppler Effect. The incident light and scattered light can be regarded as approximate parallel to the rotation axis because the skew angles are very small, when the TCs of the OAM modes are not too large. Or we can ignore the light with large skew angles because it is hard to collect. We firstly assume that the reflectivity from the spinning object is homogeneous. It means that the spinning object can be regarded as a pure phase modulator and the modulated phase depends on the roughness of surface. The roughness of stationary surface can be written as $h(r,\theta)$ shown in Fig. 4(b), so the modulated phase is given by $\Phi(r,\theta) = 4\pi h(r,\theta)/\lambda$. We rewrite the modulation function in Fourier expansion form as $\exp(i\Phi(r,\theta)) = \sum A_n(r)\exp(in\theta)$, where $\sum |A_n(r)|^2 = 1$. Considering the spin, it is revised to

$$\exp(i\Phi(r,\theta-\Omega t)) = \sum A_n(r)\exp(in\theta)\exp(-in\Omega t), \qquad (4)$$

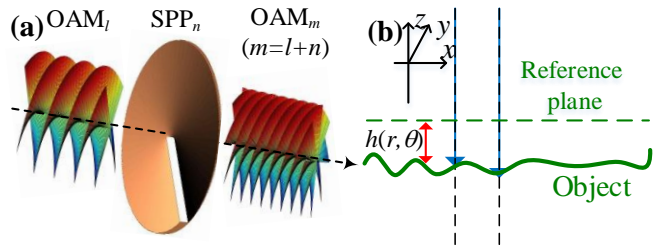

Fig. 4. (a) OAM mode conversion with a spiral phase plate. (b) Roughness of surface.

where $n$ are integers. So when an OAM mode expressed by $B(r)\exp(-i2\pi ft)\exp(il\theta)$ is normally incident on the spinning object, the scattered light can be expressed as

$$\sum B(r)A_n(r)\exp(-i2\pi ft)\exp(-in\Omega t)\exp(in\theta)\exp(il\theta). \qquad (5)$$

From Eq. (5), we can see that the scattered light is composed of many OAM modes with TCs equal to $m = n+l$, and every OAM mode has a different reduced frequency shift $(l-m)\Omega/2\pi$. These results conform to the previous analyses shown in Fig. 2 and Eq. (2). We can also see that the frequency shift content are only determined by the surface of spinning object and are not associated with the incident light. It means that as long as the surface of spinning object contains any helical phase component $\exp(in\theta)$ ($n \neq 0$), there will be frequency shifts whatever the mode content of incident light are. The incident light only determines the mode content of scattered light. So the non-twisted input OAM light (TC =0) can also have a frequency shift. These theoretical predictions are consistent with the Marrucci's report [3]. In addition, we can get the mode mapping coefficient as

$$a_{lm} = A_n(r)\exp(in\theta)\exp(-in\Omega t), n = m - l. \qquad (6)$$

More generally, the reflection from the spinning object is not homogeneous. So there are both amplitude modulation and phase modulation for the input light. The modulation function can be revised to $A(r,\theta)\exp(i\Phi(r,\theta))$, where $A(r,\theta)$ is corresponding to the amplitude modulation. It can also be rewritten in Fourier expansion form similar to Eq. (4), in this case, $\sum |A_n(r)|^2 < 1$. Then, we can expand the incident light in Fourier form based on the orthogonal OAM modes and decompose the spinning object into many SPPs, thus we can directly obtain the scattered light in Fourier form. From the expression of the scattered light, we can clearly see the frequency shift content and the process of OAM mode conversion.

This model can perfectly explain the rotational Doppler Effect. Taking the case, where two OAM modes are employed to detect the rotating body, as an example. For simplicity, we assume that only the OAM mode with $TC = m$ is selected out and the two input OAM modes are same except opposite values of TCs. So the input light can be written as

$$E_{in} = B(r)\exp(-i2\pi tf)\left[\exp(il\theta) + \exp(-il\theta)\right]. \qquad (7)$$

Then the scattered light is deduced as

$$\begin{aligned} E_{out} &= B(r)\exp(-i2\pi tf) \\ &\sum_{m \in N}\begin{cases} A_{m-l}(r)\exp(im\theta)\exp[-i(m-l)\Omega t] + \\ A_{m+l}(r)\exp(im\theta)\exp[-i(m+l)\Omega t] \end{cases} \\ &= \sum_{m \in N} C_m(r)\exp(im\theta) \end{aligned} \qquad (8)$$

Here, we assume that only one OAM mode (assuming $TC = m$) is selected out by spatial filtering and then is collected by a photodetector. Because the power of every OAM mode will not change after transmitting in free space, the collected intensity dependent on the time can be derived as

$$\begin{aligned} I_m(t) &= \iint |C_m(r)|^2 r dr d\theta \\ &\iint \begin{cases} |A_{m-l}(r)|^2 + |A_{m+l}(r)|^2 + \\ 2|A_{m-l}(r)A_{m+l}(r)|\cos(2l\Omega t + \Delta\phi) \end{cases} |B(r)|^2 r dr d\theta \end{aligned}, \qquad (9)$$

where $\Delta\phi$ is the phase of $A_{m+l}(r)A_{m-l}^*(r)$.

From Eq. (9), we can see that the OAM power has been successfully converted into the beating signal. The beating frequency is always equal to $2l\Omega/2\pi$, which does not depend on the selected OAM mode ($TC = m$). Hence, even if the detection is multimodal, all of the detected modes experience the same modulation frequency. This theoretical result agrees well with the experimental report [4]. Although the beating frequency has no relationship with the roughness of the surface, the amplitude is strongly dependent on it. Especially, if $A_{m+l}(r)A_{m-l}(r)=0$, there will be no beating frequency in the selected OAM mode. So if we want to get a highly efficient beating signal, the spinning object need carefully design. From Eq. (9), we find that the largest conversion efficiency (defined as the average power of beating signal in the selected mode divided by the total average power in all modes) is always less than 1/3 owing to the existence of direct current component and other modes. The maximum efficiency only appears under the special conditions, where $|A_{m+l}(r)|=|A_{m-l}(r)|$ and the spinning object only contains two components ($n=m-l, n=m+l$). In this case, the maximum efficiency appears only when the Fourier transform coefficients are

$$|A_{m+l}(r)|=|A_{m-l}(r)|=1/\sqrt{2} \tag{10}$$

In order to get the conversion efficiency of beating signal in OAM mode with $TC=m$, we calculate the total intensity as

$$I(t)=\iint |E_{out}|^2 rdrd\theta = [2+\cos(2l\Omega t+\Delta\phi)]\iint |B(r)|^2 rdrd\theta. \tag{11}$$

From Eq. (11), the maximum efficiency equal to 1/9 is obtained. In fact, a pure phase modulation device cannot ideally achieve the modulation function whose coefficients are expressed by Eq. (10), but it can be approximately realized by using the iterative algorithm [26, 27]. So we can optimize the parameters based on this model for some applications of rotational Doppler Effect.

This model is easily understandable and makes us better understand the physical processes of rotational Doppler Effect. It can provide theoretical guidance for many applications, such as detection of rotating bodies, detection of OAM and design of $n$ times Doppler shifter (i.e., $n\Omega/2\pi$) with a high efficiency. Let us take $n$ times Doppler shifter as an example. For a surface without careful design, the conversion efficiency of Doppler shift by a specified amount is very low. If we want to shift the light's frequency in the order of kHz with a high conversion efficiency, we need set $n\Omega/2\pi$ equal the target values and make $|A_n(r)|\approx 1$, i.e., the modulated phase of spinning surface is close to $\exp(in\theta)\exp(-in\Omega t)$. In fact, an SPP can implement this function perfectly. We just need employ an SPP with modulated phase close to $\exp(in\theta)$ as the spinning object and let the light normally passes through the spinning SPP, then the light's frequency is shifted by $n\Omega/2\pi$. In general, the rotational speed $\Omega$ can reach up to 500 rad/s [4, 5], and $n$ can also be a large value up to 40 which is decided by the fabrication technique of state-of-arts [26, 27]. So it can achieve a given frequency shift in several kHz with a very high conversion efficiency. Likewise, the conversion efficiency is only decided by the accuracy of fabricated SPP.

In conclusion, we theoretically investigate the optical rotational Doppler Effect by using modal expansion method. We find that the frequency shift content is only determined by the surface of spinning object and the reduced Doppler shift is linear to the change of OAM mode index. The theoretical model makes us better understand the physical processes of rotational Doppler Effect. We can optimize the parameters based on this model to get better performance for many applications of rotational Doppler Effect. It can provide theoretical guidance for many related applications, such as detection of rotating bodies, detection of OAM and frequency shifter.

This work was partially supported by the National Basic Research Program of China (Grant No. 2011CB301704), the Program for New Century Excellent Talents in Ministry of Education of China (Grant No. NCET-11-0168), a Foundation for the Author of National Excellent Doctoral Dissertation of China (Grant No. 201139), the National Natural Science

Foundation of China (Grant No. 11174096 , 11374008 and 61475052), Foundation for Innovative Research Groups of the Natural Science Foundation of Hubei Province (Grant No. 2014CFA004).